\documentclass[12pt]{article}
\usepackage{amsmath}
\usepackage{amsfonts}
\usepackage{enumerate}
\usepackage[title]{appendix}
\usepackage{bm,bbm}
\usepackage{mathtools}
\usepackage[authoryear, round]{natbib}
\usepackage{amssymb}
\usepackage{amsmath}
\usepackage{thmtools}
\usepackage{comment}
\usepackage{multirow}
\usepackage{array}
\usepackage{color,xcolor}
\RequirePackage[colorlinks,citecolor=blue,urlcolor=blue]{hyperref}

\usepackage{amsthm}
\usepackage[linesnumbered,ruled,vlined]{algorithm2e}

\usepackage{array}
\newcommand{\PreserveBackslash}[1]{\let\temp=\\#1\let\\=\temp}
\newcolumntype{C}[1]{>{\PreserveBackslash\centering}p{#1}}
\usepackage{colortbl}

\long\def\symbolfootnote[#1]#2{\begingroup%

\def\thefootnote{\fnsymbol{footnote}}\footnote[#1]{#2}\endgroup}

\theoremstyle{plain}

\newtheorem{theorem}{Theorem}[section]

\def\squarebox#1{\hbox to #1{\hfill\vbox to #1{\vfill}}}
\def\boxit#1{\vbox{\hrule\hbox{\vrule\kern6pt
          \vbox{\kern6pt#1\kern6pt}\kern6pt\vrule}\hrule}}

\usepackage[left=1in, right=1in, top=1in,bottom=1in]{geometry}
\usepackage{graphicx}

\DeclareMathOperator*{\argmax}{argmax}
\def\0{\mathbf{0}}
\def\1{\mathbf{1}}

\SetKwInput{KwInput}{Input}                
\SetKwInput{KwOutput}{Output}              
\begin{document}
\noindent{\large\bf Deep Nonparametric Inference for Conditional Hazard Function}

\vspace{.2in}
\noindent{Wen Su$^{1}$, Kin-Yat Liu$^{2}$, Guosheng Yin$^{3}$, Jian Huang$^{4}$ and Xingqiu Zhao$^4$\\
$^1${Department of Biostatistics, City University of Hong Kong }\\
$^2${Department of Statistics, The Chinese University of Hong Kong }\\
$^3${Department of Statistics and Actuarial Science, The University of Hong Kong}\\
$^4${Department of Applied Mathematics, The Hong Kong Polytechnic University }
}



\vspace{.2in}

\noindent{\bf Abstract:}
We propose a novel deep learning approach to nonparametric statistical inference
for the conditional hazard function of survival time with right-censored data.
We  use a deep neural network (DNN)
to approximate the logarithm of a conditional hazard function given covariates  and obtain a DNN likelihood-based estimator of
the conditional hazard function. Such an estimation approach
renders model flexibility and hence relaxes structural and functional assumptions on conditional hazard or survival functions. We establish the nonasymptotic error bound and functional asymptotic normality of the proposed estimator. Subsequently, we develop new one-sample tests for
goodness-of-fit evaluation and two-sample tests for treatment comparison. Both simulation
studies and real application analysis show superior
performances of the proposed estimators and tests in comparison with existing methods.

\vspace{.2in}

\newpage
\section{Introduction}

Censored data are often encountered in many
fields such as econometrics, biology, social sciences, and clinical trials.
Right censoring is the most typical mechanism in survival analysis. For example, in the
Study to Understand Prognoses and Preferences for Outcomes and Risks of Treatment (SUPPORT) (Knaus et al., 1995),
seriously ill patients from  five tertiary care academic centers across the United States were
followed for a 180-day period during 1989--1991 and 1992--1994.
The exact survival time of subjects who died during the follow-up period was recorded,
while those who exited the study alive were treated as right-censored.
Commonly used semiparametric
models for survival data include the Cox proportional hazards model (Cox, 1972, 1975; Andersen and Gill, 1982),
the accelerated failure time (AFT) model (Buckley and James, 1979; Tsiatis, 1990; Wei et al., 1990),
and the additive hazards (AH) model (Aalen, 1980; Cox and Oakes, 1984; Lin and Ying, 1994; McKeague and Sasieni, 1994).
However, most of the existing methods impose structural and functional assumptions on conditional hazards or conditional distributions of survival time, which may lead to unreliable and even spurious inference and prediction results when the underlying model assumptions are not satisfied.
{ {For hypothesis testing in survival analysis, Fleming and Harrington (1991) and Klein and Moeschberger (2003) provided comprehensive reviews of one-sample and two-sample tests, including the Kolmogorov--Smirnov test, log-rank test, and a class of linear rank tests. These tests do not consider conditional survival functions and thus cannot be used for testing or comparing conditional survival functions.
When covariate effects are involved, most goodness-of-fit tests focus on evaluating the assumptions of the popular Cox model
(Grambsch and Therneau, 1994; May and Hosmer, 1998; Hosmer et al., 2008; and Fox and Weisberg, 2019).
}}


Rapid growth in computational power and developments of machine learning tools
open up a myriad of new possibilities for data analysis, which enable us to build
sumptuous extensions on traditional approaches.
Examples of deep learning applications include computer vision (Krizhevsky et al., 2012; Russakovsky et al., 2015),
natural language processing (Wu et al., 2016), and reinforcement learning (van Otterlo, M. V. and Wiering, M., 2012; Silver et al., 2016).
Refer to Chakraborty et al. (2017), Murdoch et al. (2019) and Rudin (2019) for details on interpretability of neural networks.
There have also been interesting
works on applications of deep learning in survival analysis. For example, Chapfuwa et al. (2018) proposed a deep adversarial learning approach to nonparametric estimation for time-to-event analysis; Katzman et al. (2018) introduced a Cox proportional hazards deep neural network
to make personalized treatment recommendations; Lee et al. (2018) utilized a deep neural network (DNN) to learn a survival distribution directly; Kvamme et al. (2019) extended the Cox model using neural networks and constructed an approximated partial log-likelihood for estimation. { More recently, Zhong et al. (2021, 2022) provided a comprehensive literature review on combining DNN and survival analysis methods, and developed DNN approaches for a partly linear Cox model and a general class of hazards models; and more references on deep survival analysis can be found therein.}

With the goal to provide a novel framework of DNN-based statistical inference for survival analysis,
the contributions of this paper are fourfold. First, we propose a novel model-free approach to estimating the log conditional hazard function and survival function using DNNs without
imposing any assumption on its form.
As a result, any potential bias resulting from model misspecification can be mitigated. It is worth highlighting that we assume a conditional hazard rate function that does not have a baseline hazard in contrast to the Cox model, which,
to the best of our knowledge,
is completely new to the literature.
Second, we { {establish the
nonasymptotic error bound and}} investigate the asymptotic properties of the proposed estimator and establish the consistency and functional asymptotic normality. Third, we construct test statistics for one-sample and two-sample tests
as well as goodness-of-fit tests, and further derive the asymptotic normality of the proposed test statistics.
{ {In particular, we develop a new goodness-of-fit test for testing nonparametric Cox models. }}
These testing methods are breakthroughs in deep survival analysis, which bridge the statistical inferences
and deep learning. Lastly, we provide comprehensive examples in simulation studies under the Cox,
AFT and AH models
in comparisons with existing methods and show the robustness of the proposed DNN method.

The remainder of this paper is organized as follows. Section 2 introduces the proposed methodology with setups for the DNNs and exhaustive details. Section 3 presents the asymptotic error bound, and in Section 4 we first establish the functional asymptotic normality and then construct  the test statistics for hypothesis testing on goodness-of-fit as well as one-sample and two-sample tests for conditional hazard functions. Simulation studies and application to the SUPPORT dataset are provided in Sections 5 and 6, respectively. Section 7 concludes with some remarks,
and proofs of theorems are relegated to the supplementary materials.

\section{Methodology}
\subsection{Deep Neural Networks}

\noindent
We provide a brief review for the use of a deep neural network (DNN) as a function approximator (Goodfellow, Bengio and Courville, 2016).
Let $L$ be a positive integer and $\boldsymbol{p}=(p_0,\ldots, p_L,p_{L+1})$ be a  sequence of
positive integers. An $(L+1)$-layer DNN with width
$\boldsymbol{p}$ is a composite function $g:\mathbb{R}^{p_0} \rightarrow \mathbb{R}^{p_{L+1}}$ recursively defined as
\begin{eqnarray}
&&g(\boldsymbol{u})=W_L g_L(\boldsymbol{u})+\nu_L,\nonumber\\
&&g_L(\boldsymbol{u})=\sigma(W_{L-1}g_{L-1}(\boldsymbol{u})+\nu_{L-1}), \ \ \ldots, \ \
g_1(\boldsymbol{u})=\sigma(W_0\boldsymbol{u}+\nu_0), \label{e2}
\end{eqnarray}
where the matrices ${W}_l \in \mathbb{R}^{p_{l+1}\times p_l}$ and vectors ${\nu}_l \in \mathbb{R}^{p_{l+1}}$
(for $l=0,\ldots, L$) are the parameters of this DNN $g$, and $\sigma$ is the component-wise activation function, i.e., $\sigma((u_1,\ldots,u_{p_l})^\top)=(\sigma(u_1),\ldots, \sigma(u_{p_l}))^\top$. This leads to
$g_l=(g_{l1},\ldots, g_{lp_l})^\top:\mathbb{R}^{p_{l-1}} \rightarrow \mathbb{R}^{p_{l}}$ for $l=0,\ldots, L$.
The most popular activation function is
called the rectified linear unit (ReLU) defined in Nair and Hinton (2010):
$
\sigma(u)=\max\{u,0\}.
$
For the DNN in (\ref{e2}), $L$ is the depth of the network and vector $\boldsymbol{p}$ is the width of each layer,
$p_0$ is the dimension of the input variable, $p_1,\ldots, p_L$ are the dimensions of the $L$ hidden layers,
and $p_{L+1}$ is the dimension of the output layer.
The matrix entries $(W_l)_{i,j}$ are the weight linking the $j$-th neuron in layer $l$ to the $i$-th neuron in layer $l+1$, and the vector entries $(\nu_l)_i$ represent a shift term associated with the $i$-th neuron in layer $l+1$.

Following Schmidt-Hieber (2020), we
consider DNNs with all parameters bounded by one.
Let $\mathbb{N}_{+}$ be the set of all positive natural numbers. Given $L\in \mathbb{N}_{+}$ and $\boldsymbol{p}\in \mathbb{N}_{+}^{L+2}$, we consider a class of DNNs:
\begin{equation}
\begin{split}
\mathcal{G}(L,\boldsymbol{p},B) = \{g: g ~\text{is a DNN with}~ (L+1)~\text{layers and width vector} ~\boldsymbol{p}~\text{such that}\\
\max\{\|W_l\|_{\infty}, \|\nu_l\|_{\infty}\}\leq 1,~\text{for all} ~l=0,\ldots, L, \|g\|_{\infty}\leq B\},
\end{split}
\end{equation}
where $\|\cdot\|_{\infty}$ denotes the sup-norm of a matrix, a vector or a function, and $B>1$.

\subsection{Estimation}

Let $T$ and $C$ denote
survival and censoring times respectively for each subject, and let
 $X$
be a $p$-dimensional covariate vector of interest.
In the presence of right censoring, the observed data
consist of $\{Z_i=(Y_i, \Delta_i, X_i); i=1, \ldots, n\}$, which are
independent copies of $Z=(Y, \Delta, X)$, with $Y=\min(T, C)$
and $\Delta=I(T\le C)$. Assume that $T$ and $C$ are independent conditional on $X$.

With no baseline hazard, the conditional hazard rate function of $T$ given $X=x$ is
directly modeled as
$\lambda(t|x)=\exp\{g(t,x)\},
$
where $g: [0,\tau]\times[0,1]^{p}\rightarrow \mathbb{R}$ is an unknown function.
The conditional survival function and density function of $T$ given $X=x$ are respectively given by
$$
S(t|x)=\exp\left\{-\int^t_0 \lambda(s|x)ds\right\} \quad \mbox{and} \quad
f(t|x)=\lambda(t|x)\exp\left\{-\int^t_0 \lambda(s|x)ds\right\}.
$$

The log-likelihood function of $g$ based on the observed data $Z=\{Y, \Delta, X\}$ is given by
$$
l(g; Z)=\log\left\{f(Y|X)^{\Delta}S(Y|X)^{1-\Delta}\right\}=\Delta g(Y,X) -\int^Y_0\exp\{g(s,X)\}ds.
$$
Let $\mathbb{P}$ and $\mathbb{P}_n$ denote the probability measure and empirical measure, respectively. The log-likelihood of $g$ based on the observed data
is given by
$
l_n(g)=   \mathbb{P}_nl(g; Z).
$
Let ${\cal G}=\mathcal{G}(L,\boldsymbol{p}, B)$  with $p_0=p+1$ and $p_{L+1}=1$. Let $g_0$ denote the log transformation of the true conditional hazard rate function of $T$ given $X=x$. Define an estimator of $g_0$ by
$
\hat g_n\in \argmax_{g\in {\cal G}}l_n(g).
$

\section{{ Nonasymptotic Error Bound}}
To establish the theoretical  properties of the proposed DNN estimator $\hat g_n$, we assume  the true function $g_0$ belongs to a H$\ddot{\rm o}$lder class of smooth functions, which is fairly broad as discussed by Schmidt-Hieber (2020), Zhong et al. (2022) and Jiao et al. (2023). Specifically, The ball of $\alpha$-H$\ddot{\rm o}$lder
functions with radius $B_0>0$ is define as
\begin{eqnarray*}
\mathcal{C}_{p+1}^{\alpha}(B_0)&=&\bigg{\{}g:[0, \tau]\times[0, 1]^p \subset \mathbb{R}^{p+1}\rightarrow \mathbb{R}: \\
&&\quad \sum_{\boldsymbol{\gamma}:|\boldsymbol{\gamma}|<\alpha}\|D^{\boldsymbol{\gamma}}g\|_{\infty}+\sum_{\boldsymbol{\gamma}:|\boldsymbol{\gamma}|=\lfloor \alpha\rfloor}\sup_{
\boldsymbol{u} \neq \boldsymbol{v}}\frac{|D^{\boldsymbol{\gamma}} g(\boldsymbol{u}) - D^{\boldsymbol{\gamma}} g(\boldsymbol{v})| }{|\boldsymbol{u}- \boldsymbol{v}|_{\infty}^{\alpha-\lfloor \alpha\rfloor}}\leq B_0 \bigg{\}},
\end{eqnarray*}
where $\lfloor \alpha\rfloor$ is the largest integer strictly smaller than $\alpha$,
$D^{\boldsymbol{\gamma}}g(\boldsymbol{u}) ={\partial^{ |\boldsymbol{\gamma}|}g(\boldsymbol{u})}/{\partial^{{\gamma_1}}u_1\cdots \partial^{{\gamma_{p+1}}}u_{p+1}}$
 with $\boldsymbol{\gamma} = (\gamma_1,\ldots, \gamma_{r+1}) \in \mathbb{N}^{r+1}$ and $|\boldsymbol{\gamma}|=\sum_{k=1}^{r+1}\gamma_k$.

To establish the theoretical properties of the proposed estimator, we need the following
regularity conditions.
\begin{itemize}
\item[(C1)] The true nonparametric function $g_0$ belongs to $\mathcal{F}=\mathcal{C}_{p+1}^{\alpha}(B_0)$.

\item[(C2)] The covariate $X$ takes values in a bounded subset of $\mathbb{R}^p$ with a bounded
density function. Without loss of generality, we assume $X\in [0,1]^p$.

\item[(C3)]  The study stops at a finite time $\tau>0$, and there exists a small positive constant $c_1$ such that
$P(\Delta=1|X)\ge c_1$.  The conditional hazard rate function of censoring time
$C$ given $X$ is uniformly bounded,  i.e.,
$\lambda_{C|X}(t|x)\le c_2$ for $t \in[0,\tau]$ and $x\in [0,1]^p$ for a positive constant $c_2$.
\end{itemize}

For any function $g \in \mathcal{F}$, define
$
\|g\|_2=\left[E\left\{g^2(T,X)\right\}\right]^{1/2}
$
and
for any functions
$g_1$, $g_2 \in \mathcal{F}$, define
$d(g_1,g_2)=\|g_1-g_2\|_{2}.$

\begin{theorem}[Nonasymptotic error bound] \label{th3.1}
Suppose that conditions (C1)--(C3) hold. For any $N, M\in \mathbb{N}_{+}$, let ${\cal G}={\cal G}(L,  {\bf p}, B)$ be the function class of ReLU multi-layer perceptrons with width  $ \|{\bf p}\|_\infty=38({\lfloor \alpha\rfloor}+1)^2p_0^{{\lfloor \alpha\rfloor}+1}N{\lfloor \log_2(8N)\rfloor}$
and depth $L=21({\lfloor \alpha\rfloor}+1)^2M{\lfloor \log_2(8M)\rfloor}$.
Let ${\cal S}=\sum^L_{l=0}p_{l+1}(p_l+1)$ denote the size of ${\cal G}$, and $U=L\prod^L_{l=0}(p_l+1)$.
Then, we have 
\begin{align}\label{Nonasymptotic}
&\mathbb{E}d^2(\hat g_n,g_0)
\le c(B)\frac{{\cal S}\log(nU)}n+c_{0} B_0^2({\lfloor \alpha\rfloor}+1)^4
p_0^{2\lfloor \alpha\rfloor+(\alpha\vee 1)}(NM)^{-4{\lfloor \alpha\rfloor}/p_0}
\end{align}
where $c_0$ is a universal constant, and $c(B)$ is a constant depending on $B$.

\end{theorem}

The theorem is essential for deriving the asymptotic results of the proposed estimators and test statistics.

\section{Asymptotic Results}
We first establish the asymptotic distribution for the estimator $\hat g_n$,
and then apply the result to hypothesis testing.

\subsection{Asymptotic Functional Normality}

\begin{theorem}[Asymptotic normality] \label{th4.1}
Suppose that conditions (C1)--(C3) hold and  {$\sqrt{n}\tau_n^2=o_p(1)$}
where $\tau_n^2$ is the right-hand side of (\ref{Nonasymptotic}).
Then, for any $h\in{\cal F}$ with $\|h\|_{\infty}\le 1$, we have
\begin{align*}
&\sqrt n\mathbb{P}\left[\int^Y_0h(s,X)\{\hat g_n(s,X)-g_0(s,X)\}\exp(g_0(s, X))ds\right]
=\sqrt n \mathbb{P}_n\psi(g_0;Z)[h]+o_p(1),
\end{align*}
where $$\psi(g;Z)[h]=\Delta h(Y,X)-\int_0^{Y}
\exp\{g(s,X) \}h(s,X)ds.$$
Therefore,
\begin{eqnarray*}\sqrt n\mathbb{P}\left[\Delta h(Y,X)\{\hat g_n(Y,X)-g_0(Y,X)\}\right]=\sqrt n\mathbb{P}\left[\int^Y_0h(s,X)\{\hat g_n(s,X)-g_0(s,X)\}e^{g_0(s, X)}ds\right]
\end{eqnarray*}
converges to a normal distribution with mean zero and variance $\sigma^2_h=E\{\psi^2(g_0;Z)[h]\}$.
\end{theorem}
This asymptotic result plays a fundamental role  for developing hypothesis testing procedures
as discussed in the next section.

\subsection{Hypothesis Testing}
\subsubsection{One-sample Tests}
To test the null hypothesis $H_0: g=g_0$,
we let $\hat g_n(t,x)$ be the proposed DNN estimator,
and define a class of test statistics as
$$
T_w=\sqrt n\mathbb{P}_n\left[\Delta W_n(Y,X)\{\hat g_n(Y,X)-g_0(Y,X)\}\right], 
$$
where $W_n$ is a weight function.

\begin{theorem}[Asymptotic distribution of the test statistic] \label{th4.2}
Suppose the conditions in Theorem \ref{th4.1}  hold, $W_n$ is a bounded weight process and
there exists $W\in {\cal F}$ with $\|W\|_\infty\le 1$ such that
$\{\mathbb{P}(W_n-W)^2\}^{1/2}=O_p(n^{-1/4})$. Then under $H_0$,
  $T_w$ converges to a normal distribution with mean zero and variance
$\sigma^2_w=E\{\psi^2(g_0;Z)[W]\}$, where $\sigma^2_w$ can be consistently estimated by
$$
\hat{\sigma}^2_w=\frac 1n\sum^n_{i=1}\left[\Delta_i W_n(Y_i,X_i)-\int_0^{Y_i}
\exp\{\hat g_n(s,X_i) \}W_n(s,X_i)ds\right]^2.
$$
\end{theorem}

{ {For the choice of a weight function,   we follow the idea
in Klein and Moeschberger (2003, page 203) to consider the Fleming--Harrington (1991) weight
with parameters $(\rho,\gamma)$:
\begin{eqnarray*}
&&W^{(\rho,\gamma)}_n(t,x)=\frac 1n\sum^n_{i=1}I(Y_i\ge t)\hat S_n(t,x)^\rho(1-\hat S_n(t,x))^\gamma, \; \rho\ge 0, \gamma\ge 0,
\end{eqnarray*}
where $$\hat S_n(t,x)=\exp\left\{-\int^t_0\exp(\hat g_n(s,x))ds\right\}.$$
These weight functions satisfy the conditions
required by the theorem and will be used in the simulation and application studies.}}

For a given significance level $\alpha$, we reject $H_0$ if $|T_w/\hat\sigma_w|>z_{\alpha/2}$ where $z_{\alpha/2}$
is the $1-\alpha/2$ quantile of
the standard normal distribution. Define the power function of the test by
$\Gamma_n(g_1)=P(|T_w/\hat\sigma_w|>z_{\alpha/2}|g=g_1)$ under the alternative
hypothesis $H_1: g=g_1$.

{ {\begin{theorem}[Asymptotic power function of the test] \label{th4.3}
Consider the alternative hypothesis $H_1: g=g_1\neq g_0$, and let
$$\mu=E\{\Delta W(g_1(T,X)-g_0(T,X))\}\neq 0$$ and
$$\sigma^2=\mbox{Var}(\psi(g_1;Z)[W]+\Delta W(T,X)(g_1(T,X)-g_0(T,X))>0.$$ 
Then, we have
\begin{eqnarray*}
\Gamma_n(g_1)=1-\Phi\left(\frac{\sigma_w(g_1)z_{\alpha/2}-\sqrt n \mu}{\sigma}\right)
+\Phi\left(\frac{-\sigma_w(g_1)z_{\alpha/2}-\sqrt n \mu}{\sigma}\right)+o(1),
\end{eqnarray*}
where $\sigma^2_w(g_1)=E\{\psi^2(g_1;Z)[W]\}$.
\end{theorem}

It is easy to see that $\Gamma_n(g_0)=z_{\alpha/2}+o(1)$, and
the above theorem entails the consistency of the test.}}

\subsubsection{Two-sample Tests}
Consider two samples: $Z^{(k)}_i=(Y_i^{(k)}, \Delta_i^{(k)}, X_i^{(k)})$ with $Y_i^{(k)}=\min(T^{(k)}, C^{(k)}_i)$ and $\Delta^{(k)}_i=I(T^{(k)}_i\le C^{(k)}_i)$, $i=1, \ldots, n_k$, $k=1,2$. Suppose
the conditional hazard rate function of  $T^{(k)}$ given $X^{(k)}=x$ is
$
\lambda_k(t|x)=\exp\{g_{k}(t,x)\}, \quad k=1, 2.
$

Our goal is to test whether the conditional hazard rate functions from two groups are identical.
The null hypothesis is $H_0: g_{1}=g_{2}=g_0$.
Denote ${\cal O}_k=\{Z^{(k)}_i, i=1, \ldots, n_k\}$ and ${\cal O}=\{{\cal O}_k, k=1,2\}$.
Let $n=n_1+n_2$, and
rewrite ${\cal O}=\{Z_i, i=1, \ldots, n\}$ with $Z_i=(Y_i, \Delta_i, X_i)$, $i=1, \ldots, n$ under the null hypothesis.
Let $\hat g_k$ and $\hat g_0$ be the proposed estimators of $g_{k}$ and $g_0$ based on the data ${\cal O}_k$ from sample $k$  and the polled data ${\cal O}$, respectively. Define a class of test statistics as
$$
U_w=\sqrt n\mathbb{P}_n\left[\Delta W_n(Y,X)\{\hat g_{1}(Y,X)-\hat g_{2}(Y,X)\}\right], 
$$
where $W_n$ is a weight function.

\begin{theorem}[Asymptotic distribution of the test statistic] \label{th4.4}
Suppose the conditions in Theorem \ref{th4.2} hold.  If $\frac{n_1}{n}\rightarrow \xi$ as $n\rightarrow \infty$ where $0<\xi<1$, then under $H_0$, $U_w$   converges to a normal distribution with mean zero and
variance $\tau^2_w=\frac {1}{\xi}\sigma^2_{1,w} +\frac{1}{1-\xi}\sigma^2_{2,w} =(\frac {1}{\xi}+\frac{1}{1-\xi})\sigma^2_w $, where
$\sigma^2_{k,w}=E\{\psi^2(g_{k};Z^{(k)})[W]\}$, $\sigma^2_w=E\{\psi^2(g_0;Z)[W]\}$,
$\tau^2_w$ can be consistently estimated by
$\hat{\tau}^2_w=\frac {n}{n_1}\hat{\sigma}^2_{1,w}+\frac {n}{n_2}\hat{\sigma}^2_{2,w} $
or $
\tilde{\tau}^2_w=\frac {n^2}{n_1n_2}\hat{\sigma}^2_w
$
where $\hat{\sigma}^2_w$ and $\hat{\sigma}^2_{k,w}$ $(k=1,2)$ are defined
as in Theorem \ref{th4.2} based on the pooled sample and sample $k$, respectively.
\end{theorem}

The method can be extended to solving the hypothesis testing problem of multi-sample comparisons.
{ {To derive the asymptotic power function of the test statistic, we use $\hat\tau^2_w$ rather than $\tilde\tau^2_w$
because the asymptotic property of $\hat g_n$ based on the combined sample is unknown under the alternative hypothesis. For a given significance level $\alpha$, we reject $H_0$ if $|U_w/\hat\tau_w|>z_{\alpha/2}$. Define the power function of the test by
$\Gamma_n(g_1,g_2)=P(|U_w/\hat\tau_w|>z_{\alpha/2}|g_1\neq g_2)$.
\begin{theorem}[Asymptotic power function of the test] \label{th4.5}
Consider the alternative hypothesis $H_1: g_1\neq g_2$ with $\eta\neq 0$ where
\begin{eqnarray*}
\eta&=&\xi E[\Delta^{(1)}W(T^{(1)},X^{(1)})\{g_1(T^{(1)},X^{(1)})-g_2(T^{(1)},X^{(1)})\}]\\
&&+(1-\xi) E[\Delta^{(2)}W(T^{(2)},X^{(2)})\{g_1(T^{(2)},X^{(2)})-g_2(T^{(2)},X^{(2)})\}].
\end{eqnarray*}
For simplicity, assume the censoring times and covariates $(C^{(k)}, X^{(k)})$ for
the two samples have the same distribution. Let $f_k(t|x)$ be the conditional density function of $T^{(k)}$ given $X^{(k)}$   with $M_1\le f_k(t|x)\le M_2, (t,x)\in [0,\tau]\times [0,1]^{p}$ for constants $M_1$ and $M_2$, $k=1,2$.
Then, we have
\begin{eqnarray*}
\Gamma_n(g_1, g_2)
&=&1-\Phi\left(\frac{\sqrt{\frac {1}{\xi}\sigma^2_{1,w} +\frac{1}{1-\xi}\sigma^2_{2,w}}z_{\alpha/2}-\sqrt n \eta}{\sqrt{ \xi\sigma^2_{1,w}(g_1,g_2) +(1-\xi)\sigma^2_{2,w}(g_1,g_2)}}\right)\\
&&\quad+\Phi\left(\frac{-\sqrt{\frac {1}{\xi}\sigma^2_{1,w} +\frac{1}{1-\xi}\sigma^2_{2,w}}z_{\alpha/2}-\sqrt n \eta}{\sqrt{\xi\sigma^2_{1,w}(g_1,g_2) +(1-\xi)\sigma^2_{2,w}(g_1,g_2)}}\right)+o(1),
\end{eqnarray*}
where
$$
\sigma^2_{1,w}(g_1,g_2)=\mbox{Var}\left(\psi(g_1; Z^{(1)})[W+((1-\xi)/\xi)W_{11}]+\varphi_W(g_1-g_2; Z^{(1)})\right)
$$
and
$$
\sigma^2_{2,w}(g_1,g_2)=\mbox{Var}\left(\psi(g_2; Z^{(2)})[W+(\xi/(1-\xi))W_{22}]-\varphi_W(g_1-g_2; Z^{(2)})\right)
$$
with $W_{11}(t,x)=W(t,x)\{f_2(t|x)/f_1(t|x)\}$, $W_{22}(t,x)=W(t,x)\{f_1(t|x)/f_2(t|x)\}$, and \\$\varphi_W(g;Z)=\Delta W(Y,X)g(Y,X)$.
\end{theorem}

It is easy to see that $\Gamma_n(g_0, g_0)=z_{\alpha/2}+o(1)$, and the above theorem
entails the consistency of the test.}}

\subsubsection{Goodness-of-fit Tests}
Consider a hypothesis testing problem,
\begin{center}
$H_0$: $T$ follows model $M_0$ \quad vs. \quad $H_1$: $T$ does not follow model $M_0$.
\end{center}
{ {To construct a goodness-of-fit test, we randomly  split the sample into two sub-samples with
equal sample size $n_1=n_2=n/2$. Denote ${\cal O}=\{\Delta_i, Y_i, X_i, i=1, \ldots, n\}={\cal O}_{1}\cup {\cal O}_{2}$. let $\hat g_n(t,x)$ be the proposed DNN estimator of $g_0(t,x)$ based on sample ${\cal O}_{1}$, and let $\hat g_0(t, x)$ be an estimator of  $g_0(t, x)$ under model $M_0$ based on sample ${\cal O}_{2}$.
Define a class of test statistics as
$$
\tilde T_w=\sqrt n\mathbb{P}_n\left[\Delta W_n(Y,X)\{\hat g_n(Y,X)-\hat g_0(Y,X)\}\right] 
$$
where $W_n$ is a weight function.

\begin{theorem}[Asymptotic distribution of the test statistic] \label{th4.6}
Suppose the conditions in Theorem \ref{th4.1} hold, $W_n$ is a bounded weight process and
there exists $W\in {\cal F}$ with $\|W\|_\infty\le 1$ such that
$\{\mathbb{P}(W_n-W)^2\}^{1/2}=O_p(n^{-1/4})$. Further,  assume that $\hat g_0(t,x)$ satisfies
$$
\sqrt {n}\mathbb{P}_{n}\left[\Delta W_n(Y,X)\{\hat g_0(Y,X)-g_0(Y,X)\}\right]=\sqrt { n} \mathbb{P}_{n_2}\phi_w(g_0;Z)+o_p(1)
$$ with $E\phi_w(g_0;Z)=0$.
Then under $H_0$,
 $\tilde T_w$ converges to a normal distribution with mean zero and variance $\tilde\sigma^2_w=2(\tilde\sigma^2_{1w}+\tilde\sigma^2_{2w})$, where $\tilde\sigma^2_{1w}=E\psi^2(g_0;Z)[W]$ and
 $\tilde\sigma^2_{2w}=\phi_w^2(g_0;Z)$ can be consistently estimated by
$$
\widehat{\tilde\sigma^2_{1w}}=\frac 1 n\sum_{Z_i\in S_{1n}}\psi^2(\hat g_n;Z_i)[W_n] \quad\mbox{and} \quad
\widehat{\tilde\sigma^2_{2w}}=\frac 1 n\sum_{Z_i\in S_{2n}}\phi^2_{w_n}(\hat g_0; Z_i).
$$
\end{theorem}

Let $\hat{\tilde\sigma}_w$ denote the corresponding estimator of $\tilde\sigma^2_w$, and then
we reject $H_0$ if $|\tilde T_w/\hat{\tilde\sigma}_w|>z_{\alpha/2}$. Define the power function of the test by
$\Gamma_n(g_1)=P(|\tilde T_w/\hat{\tilde\sigma}_w|>z_{\alpha/2}|g=g_1)$.

\begin{theorem}[Asymptotic power function of the test] \label{th4.7}
Consider the alternative hypothesis $H_1:$ $T$ follows $M_1$ with $g_1\neq g_0$ and $\mu=E\{\Delta W(T,X)(g_1(T,X)-g_0(T,X))\}\neq 0$. Assume the conditions in Theorem \ref{th4.6}. Then, we have
\begin{eqnarray*}
\Gamma_n(g_1)&=&1-\Phi\left(\frac{\tilde\sigma_w z_{\alpha/2}-\sqrt n \mu}{\tilde\sigma_w(g_1)}\right)
 +\Phi\left(\frac{-\tilde\sigma_w z_{\alpha/2}-\sqrt n \mu}{\tilde\sigma_w(g_1)}\right)+o(1),
\end{eqnarray*}
where $$\tilde\sigma^2_{w}(g_1)=2E^2\psi(g_1; Z)[W]+2E\phi^2_W(g_0; Z)+\mbox{Var}\left(\Delta W(Y,X)\{g_1(Y,X)-g_0(Y,X)\}\right).$$
\end{theorem}
It is easy to see that $\Gamma_n(g_0)=z_{\alpha/2}+o(1)$, and the above theorem
entails the consistency of the test. }}

{ In addition, we provide two examples for Theorem 4.6 and extensions to nonparametric Cox models
respectively in Sections C and D of the supplementary materials.}

\section{Simulation Studies}

We present three simulation studies to evaluate the finite-sample performance of the proposed estimation procedure, one-sample tests and two-sample tests, respectively.

\subsection{Evaluation of DNN Estimator}
To evaluate the finite-sample performance of the proposed estimation procedure, we generated
the data from three commonly used models in survival analysis, namely,
the Cox proportional hazards model, additive hazards model, and accelerated failure time model, respectively.
We made a comparison of the estimation results from the proposed deep survival model
with those from the aforementioned models.

Suppose that covariates $X_1,\ldots,X_5$ were generated independently from a uniform distribution on $[-1, 1]$. To simulate
right-censored data, we generated censoring times $C_i$ independently from an exponential distribution with mean parameter $\mu$. We considered two censoring rates, $40\%$ and $60\%$, by adjusting the value of $\mu$.
In each simulation case, we performed $200$ independent runs with sample sizes $n=2000$ and $4000$
respectively. We tuned the number of layers, the number of neurons in each layer, and the learning rates as the hyperparameters. To obtain the optimal hyperparameter setup, we used $64\%$ of the simulated data in each simulation run as the training samples and $16\%$ of the data as
the validation samples. The rest of the data (i.e. $20\%$ of the simulated data) were used as
the test samples for the performance evaluation.

The maximization of the log-likelihood was performed by
the Adam optimizer because of its reliability among numerous models and datasets
(Kingma and Ba, 2014).
Regarding the initial values for the stochastic optimization algorithm, we adopted Pytorch's default random initialization. We used the same number of neurons in every hidden layer for simplicity, and we applied early stopping (Goodfellow, Bengio and Courville, 2016) to avoid
the overfitting problem. Specifically, we trained the neural network until the log-likelihood stopped increasing on the validation set.

To demonstrate the advantage of the proposed method, we computed the DNN estimates of conditional
hazard functions based on the simulated data generated by different survival models in comparison with
those obtained by the classical estimation approaches.

Specifically, for the Cox proportional hazards model, we considered
both linear and nonlinear covariate components:

\noindent
\textbf{Cox I: } $ \lambda(t|x) = 0.1 (t+0.01) \exp\{(x_1 + x_2 + x_3 + x_4 + x_5)/20\}$;  

\noindent
\textbf{Cox II: } $\lambda(t|x) = 0.1 (t+0.01) \exp\{(2x^2_1+ 4x_2^2 + 2x^3_3 + \sqrt{x_4+1} \log(x_5+1))\}$. 


\vspace{.1in}
For the additive hazards (AH) model, we considered

\noindent
\textbf{AH I: } $\lambda(t|x) = 0.1 (t+0.01) + (-x_1 + x_2 - x_3 + x_4 - x_5 + 15)/30$; 


\noindent
\textbf{AH II: } $\lambda(t|x) = 0.1 (t+0.01) + |-x^2_1+ 2x_2^2 - x^3_3 + \sqrt{x_4+1}  \log(x_5+1)|/2$. 

\vspace{.1in}
For the accelerated failure time (AFT) model, we considered

\noindent
\textbf{AFT I: } $\log(T) = (X_1 + X_2 + X_3 + X_4 + X_5)/20+ \varepsilon, \quad \varepsilon\sim N(0,1)$; 


\noindent \textbf{AFT II: } $ \log(T) = \cos((X_1^2+ 2X_2^2 + X_3^3 + \sqrt{X_4+1} \log(X_5+1))/20) + \varepsilon, \quad \varepsilon\sim N(0,1)$.


\vspace{.1in}
To evaluate the performance of the proposed model-free DNN approach, we examined the
discrepancy metric of the conditional cumulative hazard function (CHF):
$$
{\sum^n_{i=1}\Delta_i|\Lambda(T_i|X_i) - \hat{\Lambda}(T_i|X_i)|}\Big/{\sum^n_{i=1}\Delta_i}
$$
where $\Lambda(t|x)$ is conditional CHF of $T$ given $X=x$ and $\hat{\Lambda}(t|x)$ is the corresponding estimate.
In all cases, we compared the estimated conditional CHFs under the DNNs and
those under the traditional survival models.

For the datasets generated from Cox models I and II, we computed the proposed DNN estimates and those by fitting the standard Cox
proportional hazards model:
$$
\lambda(t|x) = \lambda_0(t) \exp(x_1\beta_1 + x_2\beta_2 + x_3\beta_3 + x_4\beta_4 + x_5\beta_5).
$$
In particular, we maximized the partial likelihood function and then
the estimate of the baseline CHF was obtained by the Breslow estimator.
For the datasets generated from AH models I and II, we computed the proposed
DNN estimates and those by fitting the standard
additive hazards model:
$$
\lambda(t|x) = \lambda_0(t)  + x_1\beta_1 + x_2\beta_2 + x_3\beta_3 + x_4\beta_4 + x_5\beta_5.
$$
We adopted the estimating equation approach and then the baseline CHF was estimated by
the
method in Lin and Ying (1994).
For the datasets generated from AFT models I and II, we computed the proposed DNN estimates and those by fitting the standard AFT model:
$$
\log(T) = x_1\beta_1 + x_2\beta_2 + x_3\beta_3 + x_4\beta_4 + x_5\beta_5 + \varepsilon,
$$
where $\varepsilon$ followed a normal distribution.
We maximized the full likelihood function under the AFT model by
assuming that $\epsilon$ followed a normal distribution.

Tables 1--3 summarize the estimation performance based on the CHF metric over  200 simulation runs.
Figures 1--3 display the pointwise averages of the estimated conditional CHFs by the proposed DNN method and the standard approaches at various values of $x$. The results suggest that the estimation performance improves when the sample size increases given the same censoring rate.
In general, the performance of the proposed method surpasses the traditional ones when nonlinearity
exists in the data generation process. These results demonstrate the robustness of the proposed DNN estimation procedure.

\subsection{Evaluation of One-sample Test}
To evaluate the performance of the one-sample test proposed in Section 4.2.1 under finite-sample situations, we considered three data generation scenarios:

\vspace{.1in}
\noindent
\textbf{Cox: } $\lambda(t|x) = 0.1 (t+0.01) \exp((x_1 + 2x_2 + 3x_3 + 4x_4 + 5x_5)/100)$; 

\vspace{.1in}
\noindent
\textbf{AH: } $\lambda(t|x) = 0.1 (t+0.01) + (x_1 + 2x_2 + 3x_3 + 4x_4 + 5x_5)/100$; 

 \vspace{.1in} \noindent
\textbf{AFT:} $\log(T)=(X_1 + 2X_2 + 3X_3 + 4X_4 + 5X_5)/5+\varepsilon$,
 where $\varepsilon$ is a standard Gumbel random variable.
%
%

\vspace{.1in}
The covariates $X_1, \ldots,X_5$ were generated independently from a uniform distribution $[0, 1]$. To simulate the right-censoring
time, $C$ was generated independently from an exponential distribution with mean parameter $\mu$.
The value of $\mu$ was chosen to yield a
censoring rate of $40\%$,  and the sample size was set as $5000$.

{  To obtain optimal hyperparameters for the estimation of $g(t,x)$, we used $80\%$ of the simulated data in each simulation run as the training samples and $20\%$ of the data as the validation samples.
The test statistic was calculated using the training data of size $n=4000$.}

For each setup, we calculated the proposed test statistic $T_w/\hat\sigma_w$ with  weights $W_n^{(\rho, \gamma)}$ given in Section 4.2, which
asymptotically follows the standard normal distribution
under the null hypothesis $H_0$. Here, we considered Cox, AH and AFT as the true model, respectively.

The test rejects $H_0$ if the test statistic $|T_w/\hat\sigma_w|$ is greater than 1.96.
Table 4 reports the estimated size and power of the proposed one-sample test over 200 simulation runs.
The estimated test sizes are close to 5\%, while the estimated power values are reasonable for all situations considered.
This result indicates that the proposed test is powerful.

\subsection{Evaluation of Two-sample Test}
To evaluate the performance of the two-sample test proposed in Section 4.2.2 under finite-sample situations, we considered three data generation scenarios:

\noindent
\textbf{Cox: } $h(t|x) = 0.1 (t+0.01) \exp((x_1 + 2x_2 + 3x_3 + 4x_4 + 5x_5)/100)$; 

\noindent
\textbf{AH: } $h(t|x) = 0.1 (t+0.01) + (x_1 + 2x_2 + 3x_3 + 4x_4 + 5x_5)/100$; 

\noindent
\textbf{AFT:} $ \log(T)= (X_1 + 2X_2 + 3X_3 + 4X_4 + 5X_5)/5 +\varepsilon$,
where $\varepsilon$ is a standard Gumbel random variable.

The covariates $X_1, \ldots,X_5$ were generated independently from a uniform distribution $[0, 1]$. To simulate
the right-censoring time, $C$ was independently generated from an exponential distribution with mean parameter $\mu$. The value of $\mu$ was chosen to yield a censoring rate of $40\%$.
During each iteration, two datasets were simulated
and the size of each dataset was set as $5000$.

To obtain optimal hyperparameters for the estimation of $g_1(t,x)$ and $g_2(t,x)$, we used $80\%$ of the simulated data in each simulation run as the training samples and $20\%$ of the data as the validation samples. The test statistic was calculated
using the training data only. Specifically, $n_1$ and $n_2$ were both set equal to $4000$.

Under each setup, we generated two datasets: One was based on $g_1(t,x)=\log(\lambda(t|x))$, while the other was based on $g_2(t,x)=\log(\lambda(t|x)) + c$ for the Cox and AFT models and $g_2(t,x)=\log(\lambda(t|x)) + c/5$ for the AH model
where $c$ represents the deviation from the null.
We obtained the size of the test by setting $c=0$ and the power of the test by setting $c=0.125, 0.25, 0.5$,
respectively.

According to the theoretical result in Theorem 4.3, we calculated $U_w/\hat\tau_w$ and the proportion of rejecting $H_0: g_1=g_2$ over 200 simulation runs. 
Table 5 reports the estimated size and power of the two-sample test.
When data were generated according to $H_0$, the rejection rate was around $5\%$;
otherwise, the rejection rate gradually increased when $g_1(t,x)$ and $g_2(t,x)$ differed farther away.

{ 
\subsection{Evaluation of Goodness-of-fit Test}
To evaluate the performance of the goodness-of-fit (GoF) test proposed in Sections 4.2.3
under finite-sample situations, we consider data generated from the following two Cox models and four non-Cox models:

\noindent
\textbf{Cox I: } $ \lambda(t|x) = 0.1 (t+1) \exp\{(x_1 + x_2 + x_3 + x_4 + x_5)/20\}$;  

\noindent
\textbf{Cox II: } $\lambda(t|x) = 0.1 (t+1) \exp\{(2x^2_1+ 4x_2^2 + 2x^3_3 + 4\sqrt{x_4+1} \log(x_5+1))/20\}$; 

\noindent
\textbf{AH I: } $\lambda(t|x) = 0.1 (t+1) + (-x_1 + x_2 - x_3 + x_4 - x_5 + 15)/30$;

\noindent
\textbf{AH II: } $\lambda(t|x) = 0.1 (t+1) + |-x^2_1+ 2x_2^2 - x^3_3 + \sqrt{x_4+1}  \log(x_5+1)|/2$. 

\noindent
\textbf{AFT I: } $\log(T) = (X_1 + X_2 + X_3 + X_4 + X_5)/20+ \varepsilon, \quad \varepsilon\sim N(0,1)$;

\noindent
\textbf{AFT II: } $ \log(T) = \cos((X_1^2+ 2X_2^2 + X_3^3 + \sqrt{X_4+1} \log(X_5+1))/20) + \varepsilon, \quad \varepsilon\sim N(0,1)$.

\noindent
Under each setup, we test `$H_0$: The Cox model is true' against `$H_1$: The Cox model is not true'.
The covariates $X_1, \ldots,X_5$ were generated independently from a uniform distribution $[0, 1]$. To simulate the right-censoring
time, $C$ was generated independently from an exponential distribution with mean parameter $\mu$.  The value of $\mu$ is chosen to yield a
censoring rate of $40\%$, and the sample size was set as $5000$.

The four GoF tests we considered are:

\noindent
\textbf{GoF test I:} The Grambsch--Therneau test that evaluates the proportional hazards assumption, see Grambsch and Therneau (1994).


\noindent
\textbf{GoF test II:} The overall goodness-of-fit test for the Cox proportional hazards model, see May and Hosmer (1998).


\noindent
\textbf{GoF test III:} The DNN-based test given by Theorem 4.6 and Example C.1 (Supplementary Materials). 

\noindent
\textbf{GoF test IV:} The DNN-based test given by Theorem 4.6 and Theorem D.2 (Supplementary Materials). 

To apply the proposed GoF test III,
we used $42\%$ of the simulated data in each simulation run as the training samples to obtain the optimal hyperparameters for the estimation of $g(t,x)$ with $16\%$ of the data as the validation samples, and the remaining $42\%$ for the estimation of the linear Cox model. To apply the proposed GoF test IV,
we used $42\%$ of the simulated data in each simulation run as the training samples to obtain $\hat g_n(t,x)$ with $8\%$ of the data as the validation samples, and $42\%$ of the simulated data in each simulation run as the training samples to obtain $\hat g_0(t,x)$ for the estimation of the nonparametric Cox model with $8\%$ of the data as validation samples.
Specifically, we set $n_1=n_2=2100$, and the test statistic was calculated using the training data only.

Table 6 reports the estimated size and power of each GoF test over 200 simulation runs. In GoF tests
III and IV, the numbers of knots were chosen according to $K_n=O(n^v)$, which were equal to 5.

The test sizes of our proposed GoF test IV were close to 5\% when data were generated from Cox models I or II, while the
tests possessed substantial power when data were not generated from Cox models. This indicates that
GoF test IV is able to differentiate Cox models from other survival models even when $g(x)$ is nonlinear.

GoF test III also yielded similar test size and power except for the case with
data generated under Cox II. This indicates that GoF test III is able to differentiate
linear Cox models from other survival models, including nonlinear Cox models.

While GoF tests I and II could produce reasonable test sizes close to 5\%, the power
values of these methods were small when data were generated from other survival models,
suggesting that classical methods might not be ideal in identifying suitable models.
}

\section{Application}
We applied the proposed method to analyze data collected from the motivating SUPPORT study.
The goal of this multi-center study was to develop an effective prognostic model to assist clinical decision making
by examining health outcomes of seriously ill hospitalized patients.
The study enrolled 9,105 patients from five tertiary care centers across the United States and was carried out in two phases,
1989--1991 and 1992--1994. The SUPPORT dataset is publicly available at \texttt{https://biostat.app.vumc.org/wiki/Main/DataSets.}
Patient data on nine diagnostic groups, physiologic variables, common laboratory measures and clinical assessment of neurologic status were recorded.
The diagnostic groups include acute respiratory failure, chronic obstructive pulmonary disease, congestive heart failure, chronic liver disease, coma, colon cancer, lung cancer, multiple organ system failure with cancer and with sepsis. Based on similarity of the survival curves, the nine diagnostic groups were aggregated into four larger disease classes: (1) acute respiratory failure (AFR) and multiple organ system failure (MOSF);
(2) chronic obstructive pulmonary disease (COPD), congestive heart failure (CHF) and cirrhosis; (3) coma; and (4) cancer.
For more details concerning the experimental design and data collection process, refer to  Knaus et al. (1995).

We considered death as the event of interest and the event time
was calculated as the number of months from study entry till death or censoring. 
We chose six covariates:  age, gender, SUPPORT physiology score (SPS), SUPPORT coma score (Scoma) and disease class. The SUPPORT physiology score was calculated based on 11 physiologic variables measured on day 3 upon entering the study,
which can serve as an overall health measure of the patient.
Refer to Knaus et al. (1995) for detailed calculation of the physiology score and coma score.
The categorical variables were converted into dummy binary variables. We set female and
COPD/CHF/Cirrhosis as the baseline levels for gender and disease class, respectively.
After removing observations with missing values, 9,104 observations remained for
our analysis with 6,200 deaths, resulting a censoring rate of $31.9\%$.
We compare the performance of our proposed DNN model with the Cox proportional hazards model
and the additive hazards model. For choosing the optimal tuning parameters in DNN,
we adopted the training (64\%) and validation (36\%)
split  and the early stopping strategy to avoid overfitting.

Figure 4 displays the estimates of conditional cumulative hazard functions of patients in each disease class and gender group at the mean values of age, SPS and Scoma scores by using the proposed DNN estimation approach.
It indicates that patients diagnosed with cancer exhibited the highest risk of death, whereas those
in the ARF/MOSP class had the lowest risk.  However,
the estimated covariate effects in Table 7 
show that ARF/MOSP could reduce the risk on the hazard, which may not be true. 
Furthermore, we applied our proposed goodness-of-fit test to the Cox proportional
hazards model   and the additive hazards model. Both the Cox model and the AH model
yielded p-values $<0.001$, which indicate that both models were inappropriate for fitting
this dataset.

\section{Concluding Remarks}

Inspired by recent developments in deep learning, we have  proposed a novel nonparametric inference approach
for conditional hazard function using DNNs. In particular, we consider a hazard function without a baseline and hence do not require any structural or functional assumptions on the conditional hazard or distribution function of survival times. Such setup
renders enormous flexibility to the model and mitigates spurious results arising from model misspecification. We have rigorously investigated the asymptotic properties of the proposed estimator and established the nonasymptotic error bound,
consistency, convergence rate and functional asymptotic normality. Subsequently, we have developed nonparametric tests
to make inference on   conditional hazard functions for evaluating goodness-of-fit and conducting treatment comparison. The proposed methods show good performances consistently in all simulated scenarios with various underlying models such as the Cox, AFT, and AH models. Specifically, in the presence of nonlinearity in the hazard functions, the proposed DNN approach yields superior performance to the traditional modeling approaches.

{  To improve the power of the proposed one-sample and two-sample tests, motivated by Yang and Prentice (2010), Lin et al. (2020) and Roychoudhury et al. (2023), we may consider the robust maxCombo test based on
the Harrington and Fleming (1982) weight.
The $p$-value of the robust maxCombo test can be obtained from a
multivariate normal distribution. If it happens
$\lambda(t|x)= 0$ for some period, we can consider $g(t,x)=\log(\lambda(t|x)+\delta_0)$, where $\delta_0$ is
a fixed small positive constant.

\section*{Supplementary Materials}

The supplementary materials include the proofs of lemmas and theorems as well as two examples and extensions.

\vspace{.1in}
\noindent{\large\bf References}
\begin{description}

\item Aalen, O. O. (1980). A model for nonparametric regression
analysis of counting processes. {\it Lecture
Notes in Statistics} {\bf 2}, Springer-Verlag, New York, pp. 1--25.


\item Andersen, P. K. and Gill, R. D. (1982). Cox's regression model for counting processes: A large sample study.
{\it The Annals of Statistics} {\bf 10}, 1100--1120.


\item Buckley, J. and James, I. (1979). Linear regression with censored data. {\it Biometrika} {\bf 66}, 429--436.

\item Chakraborty, S., Tomsett, R., Raghavendra, R., Harborne, D., Alzantot, M., Cerutti, F., Srivastava,
		M., Preece, A., Julier, S., Rao, R. M. and others. (2017). Interpretability of deep learning models: A survey of results. In {\it IEEE SmartWorld}, 1--6.

\item Chapfuwa, P., Tao, C., Li, C., Page, C., Goldstein, B., Duke, L. C. and Henao, R. (2018). Adversarial time-to-event modeling. {\em Proceedings of the 35th International Conference on Machine Learning} {\bf 80}, 735--744.
\item Cox, D. R. (1972). Regression models and life-tables. {\it Journal of the Royal Statistical Society: Series B} {\bf 34}, 187--220.
\item Cox, D. R. (1975). Partial likelihood. {\it Biometrika} {\bf 62}, 269--276.
\item Cox, D. R. and Oakes, D. (1984). {\it Analysis of Survival Data.} Monographs on Statistics and Applied Probability.
Chapman and Hall.

\item Farrell, M. H., Liang, T. and Misra, S. (2021). Deep neural networks for estimation and inference. {\it Econometrica} {\bf 89}, 181--213.
\item Farrell, M. H., Liang, T. and Misra, S. (2022)
Deep learning for individual heterogeneity: An automatic inference framework.  arXiv:2010.14694v2.

\item Fleming, T. R. and Harrington, D. P. (1991). {\it Counting Processes and Survival Analysis.} John Wiley \& Sons, Inc, New York.

\item Fox, J. and Weisberg, S. (2019). {\it An R Companion to Applied Regression.} Sage, Thousand Oaks,
CA, third edition.

\item Grambsch, P. and Therneau, T. (1994). Proportional hazards tests and diagnostics based on weighted residuals.
{\it Biometrika} {\bf 81}, 515--526.

\item Goodfellow, I., Bengio, Y. and Courville, A. (2016). {\em Deep Learning}. MIT Press.

\item Han, S., Pool, J., Tran, J. and Dally, W. (2015). Learning both weights and connections for efficient neural network. In {\em Advances in Neural Information Processing Systems} 1135--1143.
\item Harrington, D. P. and Fleming, T. R. (1982). A class of rank test procedures for censored survival data. {\it Biometrika} {\bf 69}, 553--566.
\item Horowitz, J. L. (2001). Nonparametric estimation of a generalized additive model with an unknown link function. {\em Econometrica} {\bf 69}, 499--513.
\item Hosmer, Jr., D. W., Lemeshow, S. and May, S. (2008). {\it Applied Survival Analysis: Regression
Modeling of Time-to-Event Data.} Wiley, New York, second edition.

\item Hristache, M., Juditsky, A., Polaehl, J. and Spokoiny, V. (2001). Structure adaptive approach for dimension reduction.{\em The Annals of Statistics} {\bf 29}, 1537--1566.
\item Ichimura, H. (1993). Semiparametric least squares (SLS) and weighted SLS estimation of single-index model. {\em Journal of Econometrics} {\bf 58}, 71--120.

\item Jiao, Y., Shen, G., Lin, Y. and Huang, J. (2023).
Deep nonparametric regression on approximate manifolds: nonasymptotic error bounds with polynomial prefactors.
{\it The Annals of Statistics} {\bf 51}, 691--716.



\item Katzman, J.,   Shaham, U.,   Bates, J.,   Cloninger, A.,  Jiang, T. and   Kluger, Y. (2018).
DeepSurv: personalized treatment recommender system using a Cox proportional hazards deep neural network. {\it BMC
Medical Research Methodology} {\bf 18}, 24.

\item Kingma, D. P. and Ba, J. L. (2014.) Adam: A Method for Stochastic Optimization. arXiv:1412.6980

\item Klein, J. P. and Moeschberger, M. L. (2003).
{\it Survival Analysis: Techniques for Censored and Truncated Data}. 2nd ed., Springer-Verlag, New York.

\item Knaus, W. A., Harrell Jr, F. E., Lynn, J., Goldman, L., Phillips, R. S., Connors Jr, A. F., Dawson, N. V., Fulkerson Jr, W. J., Califf, R. M., Desbiens, N., Layde, P., Oye, R. K., Bellamy, P. E., Hakim, R. B., and Wagner, D. P. (1995). The support prognostic model: Objective estimates of survival for seriously ill hospitalized adults. {\em Annals of Internal Medicine} {\bf 122}, 191--203.


\item Krizhevsky, A., Sutskever, I. and Hinton, G. E. (2012). Imagenet classification with deep convolutional neural networks. In
{\it Advances in Neural Information Processing Systems} 1097--1105.

\item Kvamme, H., Borgan,  \O. and  Scheel, I. (2019).
Time-to-event prediction with neural networks and Cox regression. {\it Journal of Machine Learning Research} {\bf 20}, 1--30.


\item Lee, C., Zame, W., Yoon, J. and van der Schaar, M. (2018). DeepHit: A deep learning approach to survival analysis with competing
    risks. {\it Proceedings of the AAAI Conference on Artificial Intelligence} {\bf 32}(1). 

\item Lin, D. Y. and Ying, Z. (1994). Semiparametric analysis of the additive risk model. {\it Biometrika} {\bf 81}, 61--71.

\item Lin, R. S. Lin, J., Roychoudhury, S., Anderson, K., Hu, T. and Huang, B. (2020). Alternative analysis methods for time to event endpoints under nonproportional hazards: A comparative analysis. {\it Statistics in Biopharmaceutical Research} {\bf 12}, 187--198.


\item May, S. and Hosmer, D. W. (1998). A simplified method of calculating an overall goodness-of-fit test for the Cox proportional hazards model. {\it Lifetime Data Analysis} {\bf 4}, 109--120.

\item Mckeague, I. W. and Sasieni, P. D. (1994). A partly parametric additive risk model. {\it Biometrika} {\bf 81}, 501--514.

\item Murdoch, W. J., Singh, C., Kumbier, K., Abbasi-Asl, R. and Yu, B. (2019). Interpretable machine learning: definitions, methods, and applications. {\it arXiv preprint arXiv:1901.04592}.

\item Nair, V. and Hinton, G. E. (2010). Rectified linear units improve restricted Boltzmann machines. In {\em International Conference on Machine Learning}, 807--814.

\item Roychoudhury, S., Anderson, K., Ye, J. and Mukhopadhyay, P. (2023). Robust design and analysis of clinical trials With
nonproportional hazards: A straw man guidance from a cross-pharma working group. {\it Statistics in Biopharmaceutical Research}
{\bf 15}, 280--294.

\item Rudin, C. (2019). Stop explaining black box machine learning models for high stakes decisions and use interpretable models instead. {\it Nature Machine Intelligence} {\bf 1}, 206--215.

\item Russakovsky, O., Deng, J., Su, H., Krause, J., Satheesh, S., Ma, S., Huang, Z., Karpathy, A., Khosla, A., Bernstein, M., Berg, C. A. and Li, F. F. (2015). Imagenet large scale visual recognition challenge. {\it International Journal of Computer Vision} {\bf 115}, 211--252.

\item Schmidt-Hieber, J. (2020). Nonparametric regression using deep neural networks with ReLU activation function. {\em The Annals of Statistics} {\bf 48}, 1875--1897.



\item Silver, D., Huang, A., Maddison, J. C., Guez, A., Sifre, L., van den Driessche, G., Schrittwieser, J., Antonoglou, I., Panneershelvam,  V., Lanctot, M. et al. (2016). Mastering the game of Go with deep neural networks and tree search. {\it Nature} {\bf 529}, 484--489.
	

\item Stone, C. J. (1985). Additive regression and other nonparametric models. {\em The Annals of Statistics} {\bf 13}, 689--705.

\item Tsiatis, A. A. (1990). Estimating regression parameters using linear rank tests for censored data. {\it The Annals of
Statistics} {\bf 18}, 354--372.


\item van Otterlo, M. and Wiering, M. (2012). {\it Reinforcement Learning and Markov Decision Processes.} In Reinforcement Learning. Springer, Berlin, Heidelberg.

\item Wei, L. J., Ying, Z. and Lin, D. Y. (1990). Linear regression analysis of censored survival data based on rank
tests. {\it Biometrika} {\bf 77}, 845--851.

\item Wu, Y., Schuster, M., Chen, Z., Le, Q. V., Norouzi, M., Macherey, W., Krikun, M., Cao, Y., Gao, Q. and Macherey, K. (2016).
 Google's neural machine translation system: Bridging the gap between human and machine translation. {\it arXiv preprint arXiv:1609.08144}.

\item Yang, S. and Prentice, R. (2010). Improved logrank-type tests for survival data using adaptive weights. {\it Biometrics}
{\bf 66}, 30--38.

\item Zhong, Q., Mueller, J. W. and Wang, J.-L. (2021). Deep extended hazard models for survival analysis. {\it Advances in Neural Information Processing Systems 34 pre-proceedings}.

\item Zhong, Q., Mueller, J. W. and Wang, J.-L. (2022). Deep learning for the partially linear Cox model. {\it The Annals of Statistics} {\bf 50}, 1348--1375.

\end{description}
\newpage

\begin{table}[!h]
\centering
{\caption{Discrepancy metric of the conditional cumulative hazard function under the Cox proportional hazards model and the proposed DNN method.}} \label{theta_bias}

\vspace{.1in}
\begin{tabular}{lcccccc} \hline
 &      & \multicolumn{2}{c}{Cox}  & & \multicolumn{2}{c}{DNN} \\  \cline{3-4} \cline{6-7}
Data Model              &  $n$   & $40\%$ & $60\%$ & & $40\%$ & $60\%$  \\ \cline{1-7}
Cox I (Linear)
&2000&0.0655&0.0441&&0.1046&0.0771\\
&4000&0.0589&0.0400&&0.0835&0.0747\\ \hline Cox II (Nonlinear)
&2000&0.5327&0.5201&&0.3036&0.2555 \\
&4000&0.4620&0.4568&&0.2825&0.2427\\ \hline \end{tabular} \end{table}

\begin{table}[!h]
\centering
{\caption{Discrepancy metric of the conditional cumulative hazard function
under the additive hazards (AH) model and the proposed DNN method.}}
\label{theta_bias}

\vspace{.1in}
\begin{tabular}{lcccccc} \hline
 &      & \multicolumn{2}{c}{AH}  & & \multicolumn{2}{c}{DNN} \\  \cline{3-4} \cline{6-7}
Data Model              &  $n$   & $40\%$ & $60\%$ & & $40\%$ & $60\%$  \\ \cline{1-7}
AH I (Linear)
&2000&0.0543&0.0359&&0.0550&0.0473\\
&4000&0.0448&0.0300&&0.0414&0.0360\\ \hline
AH II (Nonlinear)
&2000&0.3120&0.3013&& 0.22711&0.1926\\
&4000&0.2406&0.2350&&0.1787&0.1577\\ \hline
\end{tabular}
\end{table}

\begin{table}[!h]
\centering
{\caption{Discrepancy metric of the conditional cumulative hazard function
under the accelerated failure time (AFT) model and the proposed DNN method.}}
\label{theta_bias}

\vspace{.1in}
\begin{tabular}{lcccccc} \hline
 &      & \multicolumn{2}{c}{AFT}  & & \multicolumn{2}{c}{DNN} \\  \cline{3-4} \cline{6-7}
Data Model              &  $n$   & $40\%$ & $60\%$ & & $40\%$ & $60\%$  \\ \cline{1-7}
AFT I (Linear)
&2000&0.0367&0.0361&&0.0648&0.0600\\
&4000&0.0341&0.0273&&0.0481&0.0480\\ \hline
AFT II (Nonlinear)
&2000&0.6102&0.6017&&0.0877&0.0746\\
&4000&0.6086&0.6001&&0.0624&0.0524\\ \hline
\end{tabular}
\end{table}

\begin{table}[!h]
\centering
\caption{The estimated  size and power of the one-sample test.}
\label{theta_bias}

\vspace{.1in}
\begin{tabular}{ccccccc} \hline
$H_0$ &  Data Generation Model   & $W^{(0,0)}_n$ & $W^{(1,0)}_n$ & $W^{(0.5,0.5)}_n$ & $W^{(1,1)}_n$ \\ \cline{1-6}
Cox
&Cox &0.055&0.050&0.050&0.040\\
&AH &1.000&1.000&1.000&1.000\\
&AFT&1.000&0.860&1.000&1.000\\ \hline
AH
&Cox &0.995&1.000&0.920&0.800\\
&AH &0.055&0.040&0.055&0.045\\
&AFT&1.000&1.000&1.000&1.000\\ \hline
AFT
&Cox &0.635&1.000&0.505&0.980\\
&AH &0.390&0.550&0.825&0.970\\
&AFT&0.040&0.055&0.060&0.060\\ \hline
\end{tabular}
\end{table}

\begin{table}[!h]
\centering
\caption{The estimated size and power of the two-sample test.}
\label{theta_bias}

\vspace{.1in}
\begin{tabular}{cccccc} \hline
&     &     \multicolumn{4}{c}{$c$ (deviation from $H_0$)}\\
\cline{3-6}
 Model              &  $W_n$   & $0$ & $0.125$ & $0.25$ & $0.5$   \\ \cline{1-6}
Cox
&$W^{(0,0)}_n$&0.055&0.547&1.000&1.000 \\
&$W^{(1,0)}_n$&0.040&0.492&1.000&1.000 \\
&$W^{(0.5,0.5)}_n$&0.040&0.523&1.000&1.000 \\
&$W^{(1,1)}_n$&0.040&0.516&1.000&1.000 \\ \hline
AH
&$W^{(0,0)}_n$&0.045&1.000&1.000&1.000 \\
&$W^{(1,0)}_n$&0.045&1.000&1.000&1.000 \\
&$W^{(0.5,0.5)}_n$&0.040&1.000&1.000&1.000 \\
&$W^{(1,1)}_n$&0.040&1.000&1.000&1.000 \\ \hline
AFT
&$W^{(0,0)}_n$&0.050&0.510&0.800&1.000 \\
&$W^{(1,0)}_n$&0.045&0.450&0.735&1.000 \\
&$W^{(0.5,0.5)}_n$&0.050&0.500&0.715&1.000 \\
&$W^{(1,1)}_n$&0.045&0.465&0.720&1.000 \\ \hline
\end{tabular}
\end{table}

\begin{table}[!h]
\centering
\caption{The estimated size and power of the four goodness-of-fit (GoF) tests.}
\label{theta_bias}

\vspace{.1in}
\begin{tabular}{lcccc} \hline
 Data              &  GoF I & GoF II & GoF III & GoF IV  \\ \cline{1-5}
Cox I &0.045&0.035&0.055&0.055\\
Cox II &0.040&0.070&0.475&0.040\\ \hline
AH I &0.040&0.050&0.310&0.370 \\
AH II &0.585&0.465&0.490&0.530\\ \hline
AFT I &0.170&0.050&0.430&0.575 \\
AFT II &0.050&0.035&0.535&0.705\\ \hline
\end{tabular}
\end{table}

\begin{figure}[h]
\center
\label{beta_example}
  \includegraphics[width=4in]{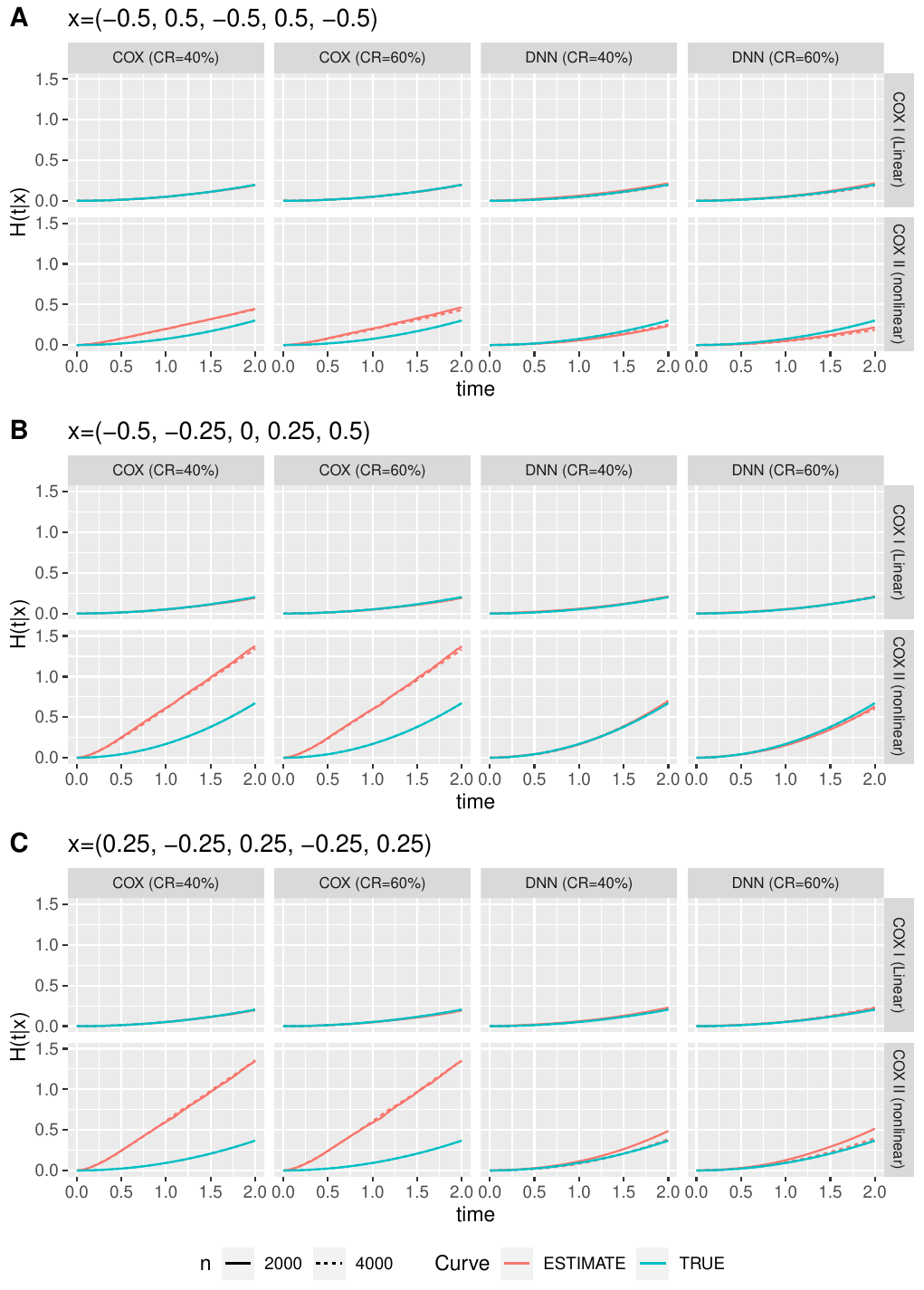}

{\caption{The pointwise averages of $\hat{H}(\cdot|x)$ estimated by the Cox model and the proposed DNN method.}}

\end{figure}

\begin{figure}[h]
\center
\label{beta_example}
  \includegraphics[width=4in]{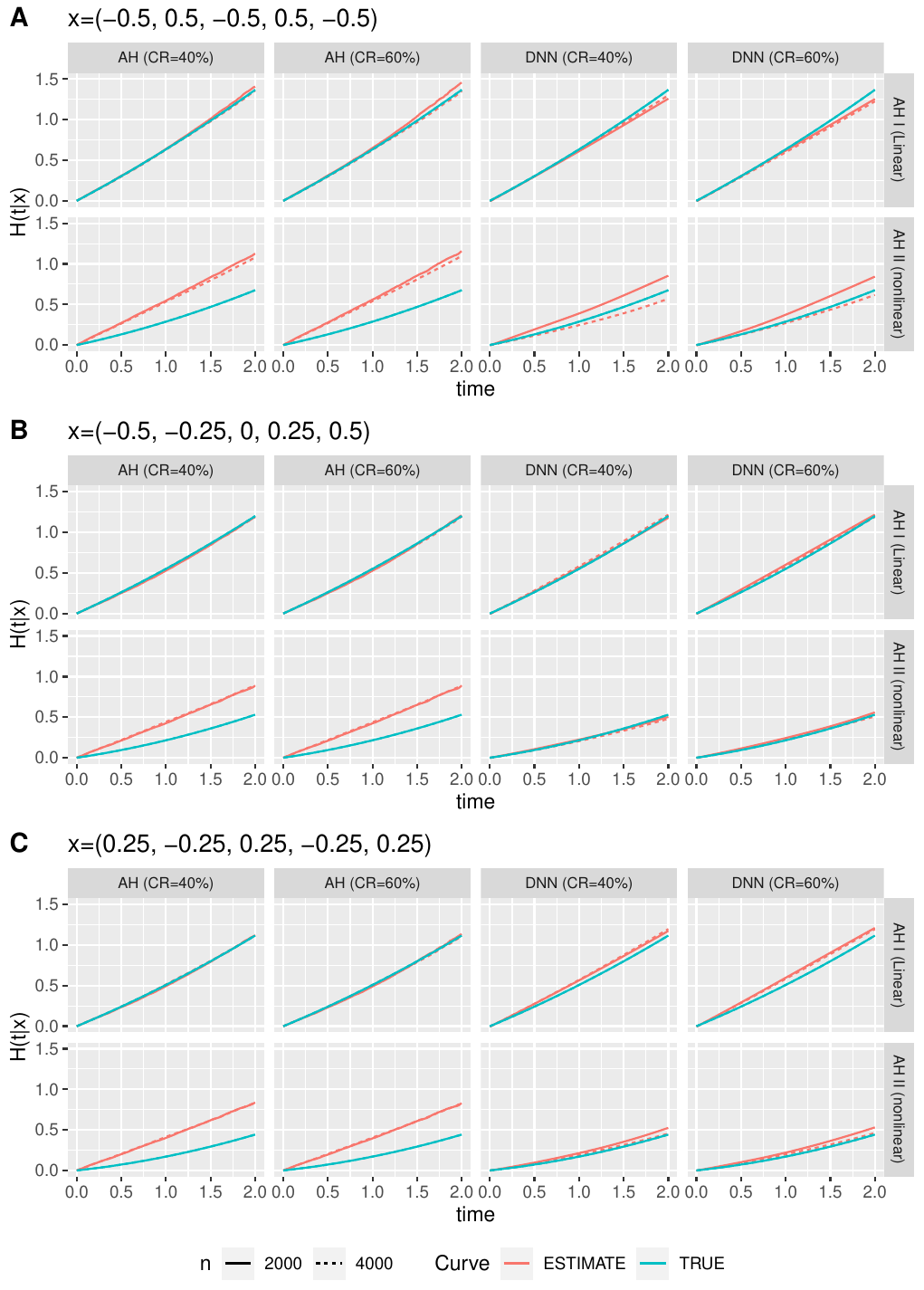}
{
 \caption{The pointwise averages of $\hat{H}(\cdot|x)$ estimated by the AH model and the proposed DNN method.}
 }
\end{figure}

\begin{figure}[h]
\center
\label{beta_example}
  \includegraphics[width=4in]{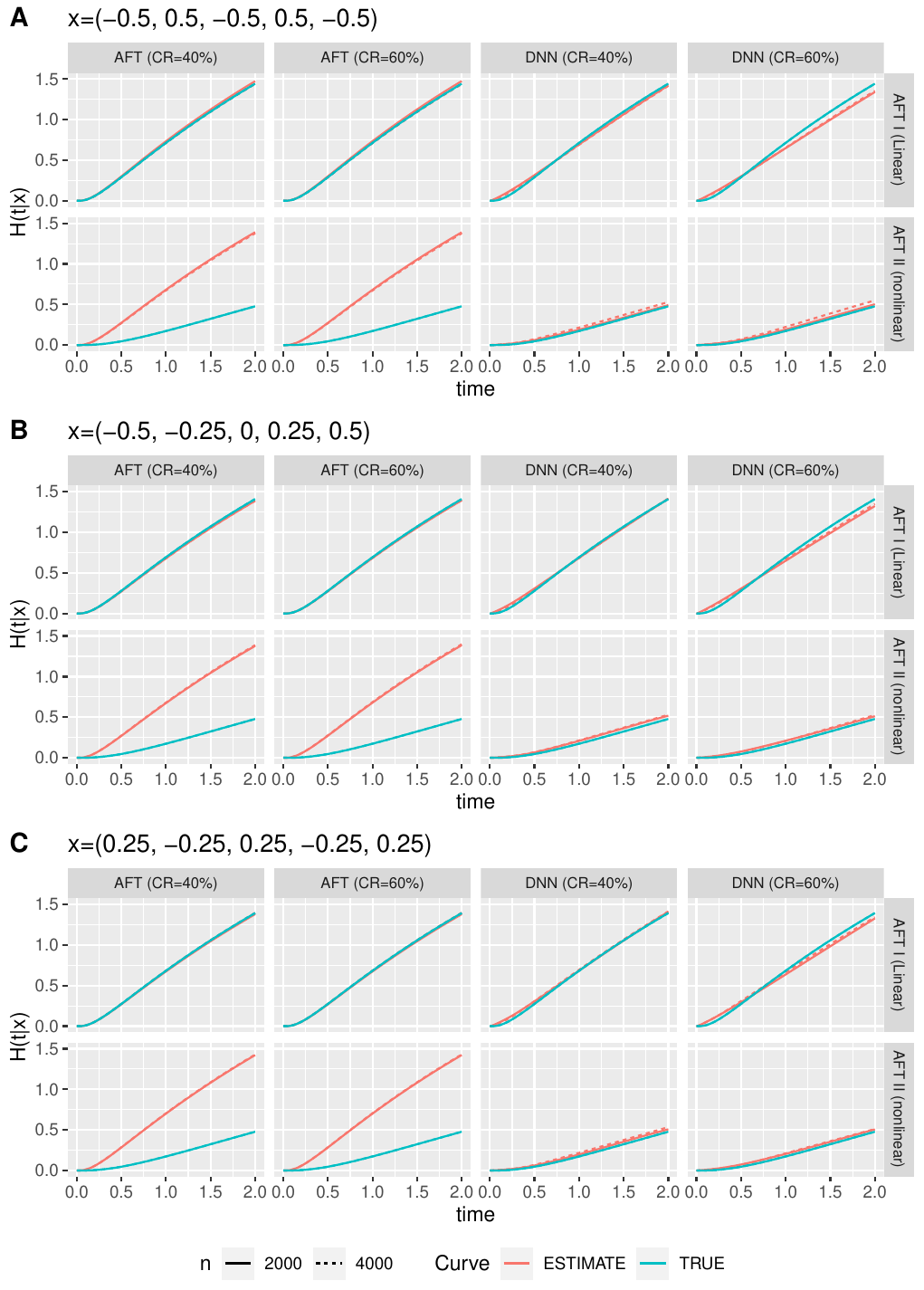}
{
 \caption{The pointwise averages of $\hat{H}(\cdot|x)$ estimated by the AFT model and the proposed DNN method.}
 }
\end{figure}

\begin{table}[!h]
\centering
\caption{The estimated coefficients and the corresponding $p$-values under the Cox proportional hazard model and additive hazards model.}

\vspace{.1in}
\begin{tabular}{lrrrrr} \hline
 &   \multicolumn{2}{c}{Cox}  & &\multicolumn{2}{c}{AH} \\  \cline{2-3}\cline{5-6}
 Covariate              &   $\hat{\beta}$ & $p$-value && $\hat{\beta}$ & $p$-value \\ \cline{1-6}
Age                     &0.0191739 &$<2\times 10^{-16}$   &&0.02766813 &$<2\times 10^{-16}$\\
Gender (Male = 1)       &0.0591388 &$0.02212$               &&0.16491274 &$5.92\times 10^{-6}$\\
SPS                     &0.0572273 &$<2\times 10^{-16}$    &&0.13776061&$<2\times 10^{-16}$\\
Scoma                   &0.0126020 &$<2\times 10^{-16}$     &&0.03714520&$<2\times 10^{-16}$\\
ARF/MOSF                &-0.1019268 &$0.00191$              &&-0.52387550&$<2\times 10^{-16}$\\
Cancer                  &0.9838260  &$<2\times 10^{-16}$              &&2.14083476&$<2\times 10^{-16}$\\
Coma                    &0.4260106  &$7.74 \times 10^{-12}$               &&0.51768792&$0.001$\\	\hline
\end{tabular}
\end{table}

\begin{figure}[h]
\center
\label{compare_icu}
  \includegraphics[width=6in]{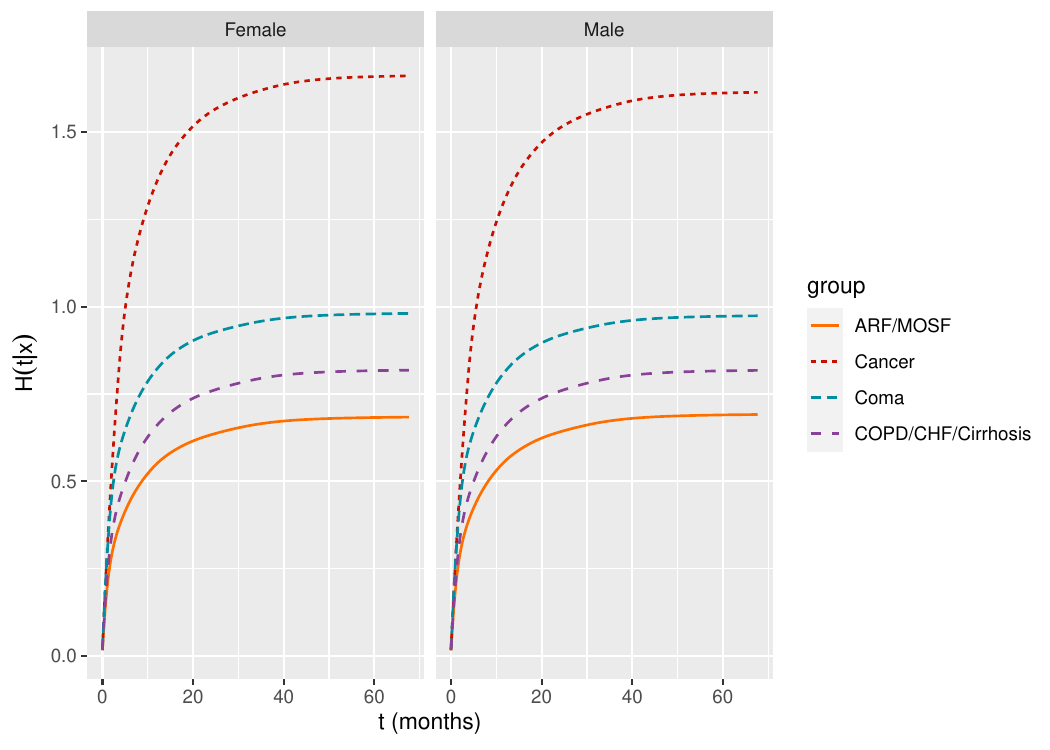}
 \caption{\footnotesize The conditional cumulative hazard functions $\hat{H}(t|x)$ of male and
 female patients in the four disease classes
 evaluated at the means of age, SPS, Scoma, which are 62.7, 25.5, and 12.1, respectively.}
\end{figure}

\end{document}